\documentclass[12pt]{article}

\usepackage{amsmath}
\usepackage{amssymb}
\usepackage{latexsym}
\usepackage{graphicx}

\def\beq{\begin{equation}}
\def\eeq{\end{equation}}
\def\bea{\begin{eqnarray}}
\def\eea{\end{eqnarray}}

\begin{document}
\begin{titlepage}

\vspace*{1cm}

\begin{center}
{\bf {\large {On the preservation of unitarity during black hole evolution 
 and information extraction from its interior}}}

\bigskip \bigskip \bigskip

{\bf Nikolaos D. Pappas}

\medskip
{\it Division of Theoretical Physics, Department of Physics,\\
University of Ioannina, Ioannina GR-45110, Greece}

\bigskip \bigskip\bigskip

\end{center}
{\bf{Summary:}} For more than 30 years the discovery that black
holes radiate like black bodies of specific temperature has
triggered a multitude of puzzling questions concerning their
nature and the fate of information that goes down the black hole
during its lifetime. The most tricky issue in what is known as
information loss paradox is the apparent violation of unitarity during
the formation/evaporation process of
black holes. A new idea is proposed based on the combination of
our knowledge on Hawking radiation as well as the Einstein-Podolsky-Rosen
phenomenon, that could resolve the paradox and spare physicists from the
unpalatable idea that unitarity can ultimately be irreversibly violated even under special
conditions.

\bigskip\bigskip\bigskip\bigskip\bigskip\bigskip\bigskip\bigskip\bigskip\bigskip\bigskip\bigskip
\bigskip\bigskip\bigskip\bigskip\bigskip\bigskip\bigskip\bigskip
\begin{center}
{ \it{E-mail: npappas@cc.uoi.gr}}
\end{center}
\end{titlepage}

\setcounter{page}{1} \noindent

\section{Introduction}

It is quite common to consider the formation/evaporation process
of a black hole as analogous to the scattering by it of particles
coming from past infinity ($\mathcal {J_{-}}$), which get measured at future
infinity ($\mathcal {J_{+}}$) by an observer living in an asymptotically flat
region of spacetime. One should note though that the term scattering is used 
in a quite extended sense. What is actually meant is that particles coming from
$\mathcal {J_{-}}$ merge and form a black hole, undergo a series of unknown
processes in its interior and, eventually, get observed at $\mathcal {J_{+}}$. 
 The simplicity of this analogy makes it a
useful tool for someone to grasp the general idea of how black
holes interact with the rest of the universe. Nevertheless, as we
have also argued in a recent work \cite{NDP}, one should always bear in mind
the limitations of this analogy, when using it. Indeed, because of our current
ignorance about the laws of quantum gravity, it is not possible to
take under consideration that, in reality, particles spend a part of their
life interacting with the singularity through these yet unknown laws
and, therefore, the aforementioned analogy is, at least, incomplete. This is
the reason why the alleged equivalence of the two processes leads to the
emergence of the deepest version of the celebrated information paradox.

More specifically, on the one hand one expects that particles,
which get absorbed by the black hole during its lifetime, are in
pure state (at least some of them, if not all). On the other hand,
Hawking radiation has been calculated to be thermal
\cite{Hawking1, HH}, that is black holes radiate like black
bodies of temperature T, directly connected to their surface
gravity $\kappa$ as
\begin{eqnarray}
T = \frac{\kappa}{2\pi}=\frac{1}{8\pi M} \quad (\rm in \; natural \; units),
\end{eqnarray}
which means that all emitted particles are in mixed state and
no correlations exist between them. All these lead us inevitably
to the conclusion that particles originally in pure
state end up in mixed state and, as a result of this evolution, a
certain amount of information about the system gets lost in an
irreversible way so it can never be recovered by any means.
However, such an evolution is not predicted in the context of
quantum physics. On the contrary, unitarity preservation, demanded
by quantum theory, requires such a case never to occur!

Encountering the whole thing as a peculiar case of scattering,
Hawking speculated that one could define an S-matrix capable of
describing the process, which he named superscattering \$ -
matrix \cite{Hawking2}. This matrix, however, should be a very
special one since it would cause the conversion of an ingoing
particle in a pure state to an outgoing particle in mixed state
and, therefore, it would be a non-unitary operator. For a matrix
like this to be allowed to exist, we should change our view of
quantum theory by introducing some new conjectures. However, such
assumptions seem to create more problems than they solve and has
been shown by the work of Banks, Susskind et al. \cite{BSP} and
Ellis, Hagelin et al. \cite{EHNS} that this cannot be the case. As
far as we know, unitarity violation that seems to occur during the
formation and the evaporation of black holes, still remains an
open issue whose answer is hoped to be found some time in the
future, after scientists have discovered and understood the nature
and the properties of the laws governing quantum gravity.

We should note here that Hawking radiation is substantially
different from the radiation emitted, e.g. by a burning piece of
coal. In the latter case, the
emitted quanta stem from the burning material itself so, once created,
they bounce off the atoms still remaining in the coal and then
carry away the information left behind in these atoms. So, at the end,
all quanta collected at infinity are entangled with themselves and
manage to carry all information existed in the initial piece of coal.
However, in the case of black holes, the quanta rise from vacuum at a
considerable spatial distance from where all matter is, therefore one
should indispensably address the question about the mechanism capable of
transferring information from the singularity to the outgoing quanta \cite{Mathur}.

We argue that any viable resolution of the apparent unitarity
violation to this problem should be based
on a combination of well known and established theories with some
innovative idea that would allow us to go a step further. The idea
proposed here is that one has to consider the
contribution of the Einstein - Podolsky - Rosen (EPR) phenomenon
\cite{EPR} in the formation/evaporation process of a black hole in
order to fully understand and efficiently describe what really
happens.

\section{Unitary scattering by the black hole}

Reflecting on the semiclassical approximation for particle
creation by a black hole at the vicinity of the horizon
\cite{Hawking1}, it becomes evident that the particles of each
pair are entangled to each other and in mixed state from the very
first moment they come into existence. While being entangled, the
particle with negative energy $E_{1}$, with respect to infinity,
propagates in the interior of the black hole all the way down to
the singularity, whereas the one with positive energy $E_{2}$ goes
away from the horizon, being the famous Hawking radiation, which
at this point is still thermal. Here comes into play the EPR
phenomenon. That is the existence of a special correlation between
particles that interacted with each other sometime and became
entangled, which holds even if the particles are separated at
infinitely large distances. Because of it each particle can
``feel'' any change in the state of the other and react
instantaneously to it \cite{EPR}.

To better understand this phenomenon, let's consider for simplicity, 
without losing any of the physics involved, a Bohmian biparticle two state system, 
where each particle can be found either in the $\mid + >$ or the $\mid - >$
state, no matter what they stand for in specific as long as
they are complementary with each other \cite{Bohm}. When
created, the wavefunction of the system is a superposition of its
two possible eigenstates, that is:
\begin{eqnarray}
\Psi = \frac{1}{\sqrt{2}} (\phi_{+}\otimes\psi_{-} -
\phi_{-}\otimes\psi_{+})
\end{eqnarray}
where $\phi_{\pm}$ and $\psi_{\pm}$ are the corresponding eigenstates
for the first and the second particle respectively. 
This description holds as long as no measurement on either particle is 
made, even though the particles can be spatially separated by large 
distances. Once an observer performs a measurement on the first particle, he/she
would find it to be in the $\phi_{+}$ or $\phi_{-}$ state. This means that
the original wavefunction has collapsed to become $\phi_{+}\otimes\psi_{-}$
or $\phi_{-}\otimes\psi_{+}$ respectively. Therefore, the observer instantly
infers with certainty that the second particle is in the $\psi_{-}$ or $\psi_{+}$ state
based on the fact that these states are (anti)correlated,
acquiring this way information about the distant particle without in
any way getting in direct contact with it.

The aforementioned standard Bohm example can be straightforwardly generalised to arbitrary
dimensional systems \cite{ArVa}. Let $ \mathcal {H}_{1} $ and $ \mathcal {H}_{2} $ be two Hilbert spaces of finite dimension
N corresponding to two subsystems $S_{1}$ and $S_{2}$ and $(\phi_{i})_{1 \leq i \leq N}$ and $(\psi_{i})_{1 \leq i \leq N}$ 
be orthonormal bases in these spaces respectively. Then for the wavefunction of the overall system we write
\begin{eqnarray}
\Psi = \frac{1}{\sqrt{N}} \sum\limits_{i=1}^N \phi_{i}\otimes\psi_{i}
\end{eqnarray}
It follows from this, as in the Bohm example, that if $\mathcal{O}_{1}$ is any observable with
N distinct values in the $S_{1}$ system, then there is an equally large observable $\mathcal{O}_{2}$ in the 
$S_{2}$ system, whose values can be predicted with certainty if we know the values of
$\mathcal{O}_{1}$ by direct measurements on the $S_{1}$ system and vice versa.

Bearing the above analysis in mind, we go back to apply it to the case of particle creation
by black holes. We shall focus our study to Schwarzschild black holes for simplicity,
since our main results remain unaltered in essence for any kind of black holes. 
We appoint the following wavefunction to the black hole
\begin{eqnarray}
\mid \Psi_{0}> = \mid n_{1},n_{2}, ... ,n_{N}>
\end{eqnarray}
where$(n_{i})_{1 \leq i \leq N}$ are the quantum numbers that correspond to the values
of the N parameters necessary to fully describe the black hole state.
Since particles emerge in pairs from vacuum in the vicinity of the horizon 
their properties have to be complementary and, therefore, their wavefunctions are expected to
be (anti)correlated. With respect to the N parameters mentioned before we write 
the wavefunctions for the first pair of particles created as   
\begin{eqnarray}
\mid 1>_{1} = \mid a^{1}_{1},a^{1}_{2}, ... ,a^{1}_{N}> \quad {\rm and} \quad \mid 2>_{1} = \mid b^{1}_{1},b^{1}_{2}, ... ,b^{1}_{N}> 
\end{eqnarray}
with $\mid 1>_{1}$ and $\mid 2>_{1}$ symbolizing the state of the first 
and the second particle of the first pair respectively and
$(a^{1}_{i})_{1 \leq i \leq N}$ and $(b^{1}_{i})_{1 \leq i \leq N}$ being the set of the values of the N
observables corresponding to them. Note that the aforementioned complementarity of the states imposes the condition  
\begin{eqnarray}
 a^{1}_{i} + b^{1}_{i} = 0  \quad\quad {\rm with} \quad\quad 1 \leq i \leq N.
\end{eqnarray}
The wave function of the pair as a whole is, of course, a superposition of all possible
eigenstates that emerge from the combination of the allowed values for $a^{1}_{i}$ and $b^{1}_{i}$.

As it is well understood, the ingoing particle (let it be the first particle of every pair) will inevitably reach
the singularity and this should happen at a finite time. To get an idea about the magnitude of the time interval required, we
use the so-called Lema\^{i}tre reference frame, which is suitable to
describe the spacetime within the Schwarzschild radius \cite{FN}. The metric in
the frame of freely falling particles has the form
\begin{eqnarray}
ds^{2}=-c^{2}dT^{2}+\frac{dR^{2}}{B}+B^{2}r_H^{2}(d\theta^{2}+sin^{2}\theta d\phi^{2})
\end{eqnarray}
where T is the proper time of the particle, $B=[\frac{3}{2r_H}(R-cT)]^{\frac{2}{3}}$
and R is the new radial coordinate. One finds, then, that the time needed
for a particle to get to the singularity starting from the vicinity of
the horizon is $t_{r_H\rightarrow0}=\frac{4GM}{3c^{3}}\sim10^{-5}\frac{M}{M_\odot}sec$. 

When the $E^{1}_{1}$-particle (the particle of the first pair with negative energy) 
arrives there, the interaction with the singularity
forces its wavefunction to collapse into one of its possible
eigenstates. It is as if the singularity performs a kind of measurement on the
ingoing particle. Then, because of the EPR-type connection between
the two entangled particles, the one with $E^{1}_{2}>0$, that has
freely propagated away from the black hole in the meantime,
instantaneously falls into the complementary eigenstate, 
therefore in a pure state. All these mean that the thermal nature of
the Hawking radiation disappears shortly after its emission. Thus, whenever it
gets to be measured by an observer living in an asymptotically flat region of
the universe, the latter would record it being in a pure state.
Finally, the black hole state becomes 
\begin{eqnarray}
\mid \Psi_{1}> = \mid n_{1}+a^{1}_1, n_{2}+a^{1}_2, ... ,  n_{N}+a^{1}_N>
\end{eqnarray}
and the second particle is found to be in an specific eigenstate
that can be measured to give us the specific values of $b^{1}_{i}$ that characterise it.

Furthermore, when the next pair of $E^{2}_1 (<0)$ and $E^{2}_2 (>0)$ particles is created,
$E^{2}_1$ falls into the black hole, reaches the singularity, that has been already modified
by its earlier interaction with the $E^{1}_1$-particle and therefore EPR-ly correlated with
the $E^{1}_2$-particle, and interacts with it modifying once more its overall state, which now
becomes 
\begin{eqnarray}
\mid \Psi_{2}> = \mid n_{1}+a^{1}_1+a^{2}_1,  n_{2}+a^{1}_2+a^{2}_2,  ... , n_{N}+a^{1}_N+a^{2}_N>
\end{eqnarray}

Then, because of the EPR phenomenon, the $E^{2}_2$-particle collapses to an eigenstate, gets correlated 
with the singularity and indirectly with the $E^{1}_2$-particle as well. 
This way all quanta emitted at early times are 
correlated with the singularity and these correlations are then transferred to the quanta
emitted at later times. This procedure continues in the same way as more and more particles
get created and emitted by the black hole. 
Therefore, after the emission of the $k-th$ particle the black hole wavefunction takes the form
\begin{eqnarray}
\mid \Psi_{k}> = \mid n_{1}+ \sum\limits_{j=1}^k a^{j}_1,  n_{2}+ \sum\limits_{j=1}^k a^{j}_2,  ... ,  n_{N}+ \sum\limits_{j=1}^k a^{j}_N >
\end{eqnarray}

At the end, all emitted quanta are correlated with each other so the
whole of the information emerges gradually as the black hole slowly evaporates. At these late 
times after having emitted an ensemble of Z particles the black hole can eventually disappear 
completely (that is $\mid \Psi_{final}> = \mid 0,0, ... ,0> $)
without any overall loss of information to occur. It is obvious that we come up with 
the following system of N equations then 
\begin{eqnarray}
n_{i} + \sum\limits_{j=1}^Z a^{j}_{i} = 0, \quad\quad 1\leq i \leq N
\end{eqnarray}
By direct measurements on every emitted
degree of freedom we get a specific set of values corresponding to all 
$(b^{j}_{i})_{1\leq i\leq N, 1\leq j \leq Z}$ parameters. Based on the (anti)correlation relation (6)
we can infer with certainty the values of the $(a^{j}_{i})_{1\leq i\leq N, 1\leq j \leq Z}$ parameters.
Then from (11) we get full knowledge of $(n_{i})_{1\leq i\leq N}$, that is the complete
set of parameters describing the initial black hole state. The whole of the information is retrieved
from black hole interior.

One could argue, though, that measurements of time scale of order
$t_{r_H\rightarrow0}$ are well within the abilities of current experiments and,
consequently, if measured at $t<t_{r_H\rightarrow0}$ the black hole radiation
will be found to be thermal and we would be confronting a non-unitary evolution
of the system. Fortunately, there is a way out in the sense that in such a case
the outgoing particles would be at a distance $r \leq \frac{5}{3}r_H$ from the singularity,
therefore, at a region that is far from being considered as flat. Unitarity, however,
results from the key demand of quantum mechanics for asymptotic completeness, which has been
established in asymptotically flat spacetime in the first place. Therefore, it is legitimate
to conjecture that in curved spacetime a deviation from unitarity could be allowed to
occur, as long as the system finally settles down in some finite time to such a situation,
that any observer at infinity would only record its overall evolution as unitary. This way
the proposed use of the EPR phenomenon, as the underlying mechanism that transfers information
from the interior of the black hole to the outgoing quanta, seems to work well in any case
without leading to paradoxes. This is a crucial assumption for the whole idea to hold, therefore, 
it needs to be further analysed. As pointed out in \cite{BirrelDavies} the scattering matrix, 
that connects the {\it in} and {\it out} states, cannot be a well-defined operator on the 
Hilbert space of states, when considering effects that take place in curved spacetimes. 
The general lack of Poincar\'{e} invariance in curved spacetimes means that there is a 
substantial bluriness in our understanding of the evolution of physical processes there. 
All these imply that there is room for a rapidly self-decaying and effectively non-observable 
deviation from unitarity to occur, when it comes to curved spacetimes.

It is also worth noting that, according to a calculation by Wald \cite{Waldbook}, observers 
near infinity should see a black hole radiate for all times $t$, such that $t-t_0>>t_D$, 
where $t_0$ denotes the time of black hole creation, as defined in \cite{Wald}, and the 
dynamical time scale $t_D$ to be
\begin{eqnarray}
t_D \sim \frac{GM}{c^3}\sim 10^{-5}\frac{M}{M_\odot} sec.
\end{eqnarray}
Therefore, in all cases there is enough time for the ingoing particle to reach the singularity, 
collapse into a pure state and provoke the corresponding transition of the outgoing particle to the 
complementary pure state, before the latter could be observed at the asymptotically flat infinity.

We should mention here that even though the explanation of the EPR
connection remains highly
controversial for over 70 years, since it was first established
theoretically, its existence is undoubted as it has also been
experimentally observed \cite{experiments}. The violation of
Bell's inequalities \cite{Bell}, that was proven beyond doubt by 
these experiments, establishes the non-local nature
of the phenomenon and it is exactly this non-locality that allows
us to postulate that the EPR connection between particles is
insensitive to the non-trivial topology of space-time near the
black hole, as to the very existence of the horizon itself.
Bearing all the above in mind, the concept of approaching the black hole
formation/evaporation as a multiparticle scattering process changes substantially. 
Particles in a pure state $\mid i >$ come from $\mathcal {J_{-}}$, merge to form a black hole, 
reemerge in the vicinity of its horizon,
get scattered and end up at $\mathcal {J_{+}}$ being in a pure state $\mid f>$. 
Therefore, at least in principle, we are able to define an
S-matrix that can describe this procedure having the very
important property to be unitary for all observers located near infinity, as it ultimately predicts the evolution of a
pure initial state into a pure final state. The matrix elements 
\begin{eqnarray}
< i \mid S \mid f >
\end{eqnarray}
can, obviously, only be determined {\it a posteriori}, should we ever be able to 
observe the complete creation and evaporation process of a black hole in a fairly 
known and controlled environment (hopefully at the LHC, if any theory predicting
the Planck scale to be as low as few TeV \cite{theories} is proven to be right.
In this case we would be able to produce a multitude of mini black holes using conveniently prepared
particles and record a detailed evaporation profile of them.
Then our measurements would provide us
with the transition amplitudes connected to this procedure and the validity of our approximation could 
be tested against the experimental results.). This limitation, however, is 
inherent in the study through scattering of every system, for which we lack a complete
microscopic description, and should not be seen as a fatal flow 
of our approach. Even though the determination of the matrix elements 
would probably be very difficult in practice (also because of the arbitrarily large number 
of dimensions that it needs to have in the case of massive black holes), the important thing 
is that this matrix can be defined in the first place and sought for.

\section{Discussion}

In the years before the unexpected and fairly shocking discovery
that black holes actually radiate, all kinds of information, that
got swallowed by them during their lifetime, were considered to
remain eternally trapped in their interior and, therefore, no
physical or philosophical problem occurred. The knowledge that
this is not the case gave rise to a serious paradox, which verge
on the very foundations of quantum physics, as it seems that a
particle in a pure state can evolve into some mixed state during
the formation/evaporation process of a black hole and,
consequently, information about the system gets lost for ever.
This is, in a nutshell, the celebrated information loss paradox
or, at least, the most difficult to address version of it. It
appears that the combination of the semiclassical approximation
for particle creation by black holes, as presented by Hawking,
with the EPR connection between entangled particles, as established by
Einstein, Podolsky and Rosen, can provide us with a viable
explanation of what really happens during the evaporation of a
black hole, without assuming that unitarity can be violated
against all requirements of quantum theory.

It is only enough the EPR phenomenon to hold also in the case of black
holes and there is no apparent reason why it shouldn't do so.
Then, all particles will evolve into some pure state soon after
their creation due to the influence of the singularity on them
both directly (by contact) and indirectly (through the EPR
correlation), so that any observer away from the black hole would
receive and measure them to find that they all are in a pure
state. In this way unitarity remains safe and no paradox occurs to
undermine the credibility of our quantum mechanical understanding
of nature.

Furthermore, this procedure allows us, as
external observers, to extract information from the black hole interior.
Every emitted particle, that gets measured, provides us with an elementary
piece of information regarding the internal states, reducing our
ignorance about them. Recalling that the entropy of a system is considered
to be directly correlated to our lack of knowledge about it, the aforementioned
gain of information about the black hole state should lead to a decrease of its
entropy. On the other hand, emission of Hawking radiation results to the shrinking
of the black hole mass and, consequently, its surface area A, which is proven to
be connected to its entropy through the famous Bekenstein - Hawking formula
$S=\frac{A}{4}$ (use of natural units, A measured 
in planckian units). We argue that this concordance of predictions about the
entropy, derived by two quite different starting points, is not incidental and
should be seen as an extra argument in favour of the validity of our analysis.

As closing remark, we think that it is reasonable to believe that the
proposed combination of the semiclassical approximation
for particle creation with the principles of the EPR phenomenon, 
the way it was analysed earlier, is valid and, in any case, we hope that this
work would be considered a useful and inspiring contribution to
the effort to achieve a deeper and more complete understanding of
black holes and their physics.

{\bf Acknowledgements:} The author is deeply grateful to Prof. Panagiota Kanti for useful
discussions and guidance through the relevant literature.

\end{document}